\documentstyle[eqsecnum,aps]{revtex}
\input psfig.tex

\def\be{\begin{equation}}
\def\ee{\end{equation}}
\def\bea{\begin{eqnarray}}
\def\eea{\end{eqnarray}}

\def\cmm2{{\,\rm cm^{-2}}}
\def\cm2{{\,{\rm cm}^2}}
\def\cmm3{{\,{\rm cm}^{-3}}}
\def\gcmm3{{\,{\rm g\,cm^{-3}}}}

\def\fun#1#2{\lower3.6pt\vbox{\baselineskip0pt\lineskip.9pt
  \ialign{$\mathsurround=0pt#1\hfil##\hfil$\crcr#2\crcr\sim\crcr}}}

\begin{document}
\draft
\preprint{DAMTP-1998-78}
\title{Exactly Azimuthal Pixelizations of the Sky }
\author{Robert G. Crittenden and Neil G. Turok}
\address{DAMTP, CPAC, University of Cambridge}
\date{\today}
\maketitle

\begin{abstract}
We investigate various pixelizations of the sky which allow
for fast spherical transforms,
for implementation in full sky CMB experiments such as Planck and MAP. 
We study the effect of varying 
pixel shape and area on the extraction of the CMB power spectrum.
We argue for the benefits of having a truly azimuthal, or `igloo' 
pixelization.  Such pixelizations are simple and 
allow
for fast, exact simulations of pixelized skies. They also
allow for precise correction to be made which accounts for
the effects of pixel smoothing on extracted multipole moments. 
\vskip 0.2in 
\end{abstract}


\section{Introduction}

The prospect of high precision all sky maps of the cosmic microwave
sky from the MAP and Planck satellites has galvanized 
cosmology. These maps will contain millions of accurate, independent
temperature measurements which can be used to test theories
of the early universe and cosmic structure formation. This paper
deals with the very practical issue of how to store the data so that
it can be easily manipulated and analyzed. In particular we discuss
the question of how to pixelize the sky. This is clearly a question of
some importance: data of this quantity and quality are likely
to form the basis for much of cosmology over the next decade. 
As we shall discuss,  
the right choice of pixelization will both dramatically improve and
accelerate analysis. 

Numerous suggestions for pixelization of the sky 
have already been made \cite{sk97,quad}.
The simplest uses 
equal divisions in latitude and longitude ($\theta$ and $\phi$).  
This has been called the
Equidistant Cylindrical Projection (ECP). It has the
advantages of 
being both azimuthal and simply hierarchical, in that the data 
can be easily coarse grained by combining neighbouring pixels.
The azimuthal symmetry allows for fast 
spherical harmonic transforms, speeding
many operations 
such as map simulation and inversion\cite{dh94,moh97,mnv97}. 

The biggest 
failure of the ECP pixelization is that the 
pixels near 
the poles are small and very
distorted. In a real experiment, they might be very noisy or even contain 
no data at all. These extra pixels are also wasteful in later analysis. 
The ECP scheme  can be improved upon by grouping more and more pixels 
together as one approaches the pole, and this is the idea 
behind the
pixelizations we advocate  below. 

Previous all sky CMB maps, produced by the 
COBE satellite, used a pixelization based on the 
Quadrilateralized Sky Cube Projection, or `quad cube' \cite{quad}.
The edges of a cube are projected onto a sphere, dividing the sky into six
equal areas. These are subdivided
into a roughly square, hierarchical lattice.
The main drawback of the resulting pixelizations is 
their lack of azimuthal symmetry, making 
spherical harmonic transforms  time consuming. 
Other suggested pixelizations have hexagonal instead of 
square pixels. One is an icosahedral version of the quad cube
\cite{teg96}, another is a class based on fullerenes
\cite{bat97}. While these pixels are closest to
the ideal, round pixels, they lack azimuthal symmetry or 
any obvious hierarchical structure for coarse graining. 

Gorski and collaborators have recently made available a package
of programs, called HEALPIX, which include a novel pixelization scheme
based on a rhombic dodecahedron and associated software for
performing fast spherical transforms
\cite{gor97}. The centers of the HEALPIX pixels are 
located on a lattice possessing discrete azimuthal symmetry at each 
latitude. Such symmetry is essential for making use of
Fast Fourier transforms with respect to the azimuthal coordinate $\phi$. 
However in HEALPIX the pixel shapes vary around circles of constant latitude,
so the integrations needed for spherical transforms are only
performed approximately, effectively assuming the pixels are 
azimuthally symmetric, identical and round.
In contrast, the pixelizations we consider do possess exact azimuthal
symmetry at each latitude, so that we can perform the 
the integrations needed for spherical transforms 
exactly.

An attractive feature of the `igloo' schemes we focus on
is that they have an exact discrete azimuthal symmetry at each latitude,
allowing fast and exact spherical harmonic transforms. In particular,
this allows one to rapidly create simulated skies in which the
effects of pixels smoothing have been exactly included (up to machine
accuracy). The ability to perform such transforms quickly is also likely
to be essential to computing the angular power spectrum of the
observed sky (for a recent discussion see \cite{osh98}). 

For comparative purposes we consider several different igloo schemes,
which satisfy the following criteria to varying extents:

\begin{itemize} 
\item Lack of Pixel Distortion -- 
Pixelization of the data suppresses modes with wavelengths shorter than
the pixel dimensions in any given direction. To minimize this effect
one should try to make the largest pixel diameter as small as possible. 
The ideal (unachievable) limit would be to have circular pixels 
where the largest diameter is $ D = 4/\sqrt{N_{tot}}$ where $N_{tot}$
is the total number of pixels on the sky. 
(One can see this by considering the average pixel area, 
$A = 4 \pi / N_{tot}  = \pi D^2/4$.) 
For square pixels, the best one can achieve is $D = \sqrt{8\pi/N_{tot}}$, 
about 25\% larger.
In general we 
define the distortion of a pixel to be the length of longest diameter
divided by this ideal value $D$.

\item Equal Area Pixels  -- 
This is desirable in order to get the best resolution 
for a fixed number of pixels, and so that the pixels,
in the first approximation, have equal statistical weight.
However, since sky coverage and and foregrounds are unlikely 
to be uniform, it is not clear that exactly equal area pixels are
essential.

\item Built-in Hierarchy -- A hierarchical pixelization scheme allows
for coarse graining of the data at different resolutions. This is
likely to be at least very useful, and probably essential, 
in analyzing the data. We require that the
hierarchy be nested, that is that each higher resolution
pixels should fit perfectly into a single lower resolution pixel. 
Ideally one would like the hierarchy to coarsen to as small a
base set of pixels as possible, maintaining minimal distortion
and roughly equal area pixels at each level. 

\end{itemize}

In section II, we describe the general features of igloo 
pixelizations as well as specific characteristics of the models that we 
consider here.  In section III, we present some general considerations 
for creating and inverting pixelized maps, and describe how the azimuthal 
symmetry can be exploited.  Finally, in section IV, we 
create and invert realistic maps with the various pixelizations and 
examine the accuracy with which the power spectra can be extracted.

\section{Igloo Tilings}
\subsection{General Description}
 
By an igloo pixelization, we mean one divided into rows 
with edges of constant latitude and where each row is 
divided into identical pixels by lines of constant longitude.
The pixels are roughly trapezoidal shaped,  
becoming nearly 
rectangular away from the poles. 
For simplicity and to speed calculations, the northern and southern 
hemispheres are tiled identically.

Igloo tilings have many advantages.  First, they are quite simple.
They are also naturally azimuthal and can be easily made 
equal area, with most pixels nearly square. 
But perhaps their biggest advantage is
that
the pixel edges are defined along constant lines of the spherical
polar coordinates $\theta$ and $\phi$, allowing for 
an exact, fast integration of spherical harmonics
over the pixels. This is essential in constructing exact simulated
skies, and in optimally recovering the sky power spectrum from
real data.

The simplest example of an igloo pixelization is the ECP 
pixelization itself, which 
we will include in our comparison.
In it, every row has the same number of pixels, and the pixels become quite 
narrow near the poles.		
The other models we consider have pixels which are 
more nearly square and nearly (or exactly) equal area. To do this
the 
number of pixels in each row must decrease as one approaches the poles
(Figure 2). 

One can construct igloo pixelizations with either 
rows equally spaced in latitude, like the ECP,  
or with pixels of uniform area. 
The advantage of equal latitude spacing is 
that the pixelization can be  
created by a simple rebinning of an ECP pixelization, providing 
the latter is chosen to have an appropriate number
of pixels. 
In addition, by letting the 
pixel areas vary, one can make them less distorted. 
An equal area pixelization will not be exactly 
equally spaced in latitude, but has the advantage that
all of the pixels will have 
the same statistical weight. 

Igloo pixelizations can also be made hierarchical. 
To do this, one first  
divides the sphere into a base 
pixelization with relatively few pixels, 
optimized to minimize the pixel distortion. 
Each of these coarse pixels is then divided into four by 
bisecting it in longitude and 
latitude.  The latter division is chosen either to keep the pixels 
the same area or to maintain a constant latitude spacing of the rows. 
Thus one 
creates a finer grained pixelization with four times the number of pixels. 
This procedure can be repeated until we reach the required resolution. 

Reducing the number of base pixels tends to increase 
the level of pixel distortions, so some compromise must be found. 
While there are clear advantages to having 
fewer base pixels, it is not obvious how they should be weighed against the 
advantages of having more uniform pixels.
Here, we will consider two possible extremes, 
pixelizations with twelve 
base pixels (roughly $60^o\times60^o$ each),  
and one with 12,116 $2^o\times2^o$ base pixels. 
The latter number was chosen because an analysis at this 
resolution is comparable to what has already been accomplished for COBE, and it
is unlikely one will need an analysis at a lower resolution.  

The hierarchical division causes some of the subpixels to become 
more distorted than the coarser pixels, especially near the poles.
Away from the poles, 
there is a limit to how distorted the pixels become, even 
at the highest resolutions.  
However, if the polar cap were simply bisected in $\theta$ and $\phi$, 
the pixels would become more and more distorted, 
as occurs in the ECP.  
Thus, we must use another method for dividing the polar regions.

\begin{figure}[htbp]
\centerline{\psfig{file=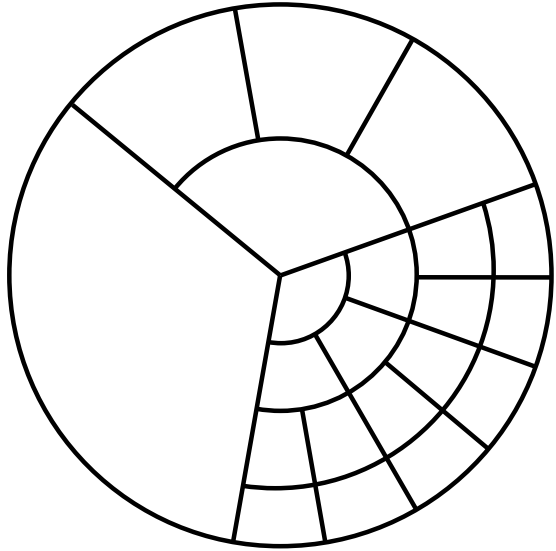,width=2.9in}}
\caption{
Picture of the polar cap region in the igloo schemes, showing 
three levels of subdivision to higher resolution pixels.
}
\label{fig:cap}
\end{figure}

In our pixelization, we have chosen to initially divide 
the cap of each pole into three equal wedges.
Higher resolution pixelizations are found by dividing each wedge into  
four pieces, one 
central wedge and three pieces surrounding it.  
(See Figure \ref{fig:cap}.)
This process is iterated, with the 
interior wedge always being divided in this way and the outer pieces 
being divided by lines of constant $\theta$ and $\phi$.   
This prescription is designed in order to 
minimize the distortions in the highest resolution pixels.  

\subsection{The Base Pixelization}

We would like to begin with 
base pixels that are as uniform and undistorted as possible.
We can make the pixels nearly square 
by choosing the 
right number of pixels in each row.
The best number at a given latitude is 
given by the distance around the sphere at that latitude
divided by the width of 
each pixel.
Thus, if the average area of each pixel is 
$A = 4\pi/N_{tot}$, then the ideal number in a row at angle $\theta$ is  
$N(\theta) = 2 \pi \sin (\theta)/ \sqrt{A} = \pi^{1/2} \sin (\theta) 
\sqrt{N_{tot}}$.    Since the number of pixels must be an integer, 
some slight distortion of the pixels is inevitable. 

If we treat the region near the pole as effectively flat, then it is simple to  
calculate the ideal numbers in the concentric layers.   
Assuming the pixels are square, the radius for each layer increases 
by $\sqrt{A}$, so that the ring area increases by $2\pi A$.
Thus, the number in a layer is related to the number in the previous layer by 
$N_p = N_{p-1} + 2 \pi$.  
If the cap is initially divided into three wedges,
the next row should contain nine pixels, and there should be 
roughly six 
more pixels in each additional row.
The precise numbers are chosen to minimize pixel distortion at the 
highest resolutions.  One possible sequence might be
3, 9, 15, 22, 28, 34, 40, 46...     

Having arbitrary integers in the rows 
reduces the effectiveness of Fast Fourier Transforms (FFTs), 
whose speed depends on 
the number of pixels in a row being a product of low primes.
To minimize this effect, one should be selective in 
choosing the number of pixels allowed in a row. 
For example,  
we may chose only numbers of the form, $2^n$ or $3^m\times2^n$. 
Since allowing more possibilities reduces the amount of pixel distortion, 
one might alternatively consider all numbers 
which do not have large prime factors.  
For example, 
in the 12,116 base pixel model we 
exclude only numbers with prime factors bigger than 23.  

We parameterize the distortion in terms of the largest pixel diameter, $D$, 
which is related to the 
pixel height, $h$, and its width, $w$:
$D^2 \simeq h^2 + w^2$.
The width is given by $w = 2\pi \sin(\theta)/N_p = \sqrt{A}N(\theta)/N_p$, 
where $N(\theta)$ is the ideal number of pixels at a given latitude and 
$N_p$ is the actual number. 
In the equal latitude pixelizations, the pixel height is fixed (ideally at 
$h = \sqrt{A}$,) so that 
\be
D^2 \simeq A (1 + (N(\theta)/N_p)^2).  
\ee
In the equal area case, the height varies inversely with the width, so 
that 
\be D^2 \simeq A ((N_p/N(\theta))^2 + (N(\theta)/N_p)^2). \ee 
The distortions are smaller on average in the constant latitude case, 
but the pixel areas vary from row to row.

If one allows large number of base pixels, then the pixel widths and heights 	
can be made almost equal ($D^2 = {2A}$) for most of the pixels, with the
distortions being the worst at the poles. 
Even in the finest subdivision however, the pixel distortions 
do not become arbitrarily bad.  
The most distorted pixels 
never have diameters larger than about $D^2 \simeq 4A$ in any of the 
pixelizations we consider, and these are relatively rare.
  
These pixelizations are optimized to be both azimuthal and hierarchical. 
The data need to be ordered differently 
depending on which of these properties we are interested in utilizing. 
For the creation and analysis of maps, 
it is helpful to have an azimuthal ordering.
However, 
to facilitate coarsening, one would reorder the data 
into chunks associated with the coarsest graining.  Within these chunks, the 
data would be ordered like each face in the quad cubed, so that adding the 
data in groups of four converts it to the next coarser level. 

\subsection{Specific Igloo Tilings}

For definiteness, here we consider four possible igloo pixelizations: 
\begin{itemize}
\item an ECP pixelization 
\item a twelve pixel scheme (3:6:3) divided with equal area
\item a twelve pixel scheme (3:6:3) divided with equal latitude spacing
\item an equal area scheme with 12,116 base pixels 
\end{itemize}

In the twelve pixel scheme, the base pixels are 
arranged as shown in Figure 2 (bold lines), with three pixels 
at each cap and six around the center.
The divisions between the layers lie at $\theta = \pm 30^\circ$.
(We note that
the pixelizations recently suggested by Gorski \cite{gor97} also
involve twelve base pixels, but arranged instead in rows of 4:4:4.)
To facilitate comparisons, the ECP model was chosen to have nearly the same 
number of pixels as the other models, and was based on a $5\times10$ grid, 
with the base pixels having edges of $\Delta \theta = \Delta \phi = 36^\circ$. 

\begin{figure}[htbp]
\centerline{\psfig{file=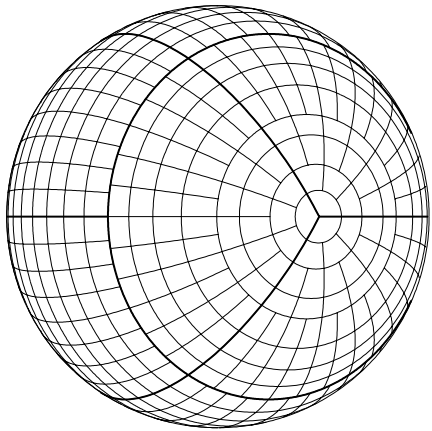,width=2.9in}}
\caption{
Picture of the 3:6:3 equal area pixelization, which divides the sky into twelve 
base patches, three at either cap and six $60^o\times60^o$ patches at the 
equator.  Here, each base pixel is broken up into 64 smaller pixels.
}
\label{fig:363}
\end{figure}

\begin{figure}[htbp]
\centerline{\psfig{file=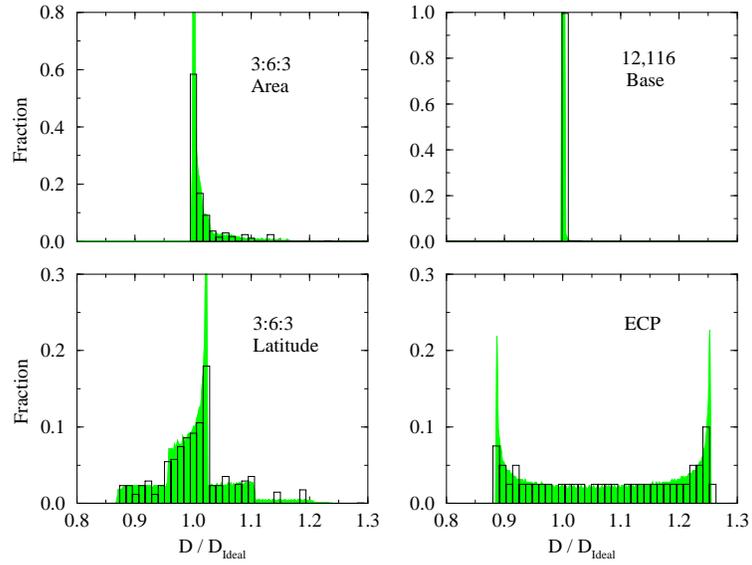,width=3.9in}}
\caption{
The percentage of pixels with a given maximum diameter for each of the 
pixelizations, normalized in terms of the ideal diameter for square pixels, 
$D_{Ideal} = \sqrt{8\pi / N_{tot}}$.   
The histograms show the distribution for $2^\circ$  
pixelizations, while the shaded region shows the results when each 
component pixel is divided into 1024 subpixels ($3.5'$ resolution.)
}
\label{fig:distort}
\end{figure}

{\bf Distortions:}
We evaluate these schemes given roughly equal numbers of pixels in each. 
We expect naively that the quality of the pixelization should 
depend on the average diameter of the pixels. 
Figure \ref{fig:distort} 
shows the percentage of pixels with a given maximum diameter
for the various models. 

The least distortions result from the pixelizations with the greatest number 
of base pixels, and for 12,116 base pixels the amount of distortions 
is very close to the ideal (square) result.  
When the number of base pixels is reduced to twelve, greater distortions 
result.  In the equal area case, the minimum diameter is fixed and grows 
larger for pixels that are less square.  
In the equal latitude case, the distortions are comparable.  
However, the pixels with less area can have a smaller diameter, so that the
average diameter is actually smaller than in the equal area case. 
Finally, the distortions in the ECP are the largest because many of its pixels 
contain very little area, which makes the remaining ones larger.  

\begin{figure}[htbp]
\centerline{\psfig{file=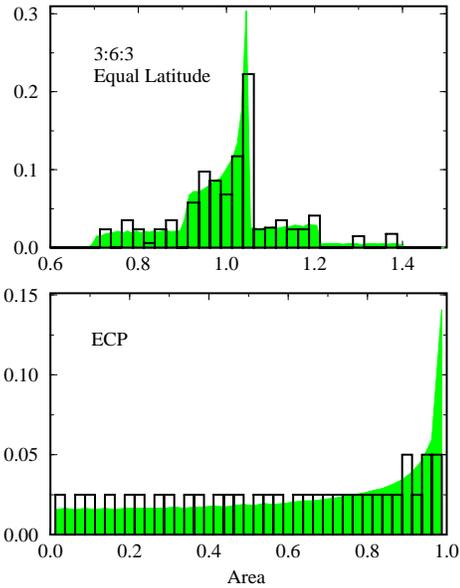,width=3.9in}}
\caption{
The distribution of the area of the pixels for the unequal area models.
The top figure is normalized so that the average area is one, while the bottom 
figure is normalized so that the area of the pixels along the equator is one. 
Again, the histograms show the distribution for $2^\circ$
pixelizations, while the shaded region shows the results when each 
component pixel is divided into 1024 subpixels.
}
\label{fig:area}
\end{figure}

{\bf Area:}
Only two of these pixelizations, the twelve pixel equal latitude and the ECP
pixelization, are not equal area by construction.
We might hope that variations in pixel area themselves will have 
little effect, because we
calculate the spherical transforms properly weighting for the different 
pixel areas. 
Nevertheless the
ECP scheme is at a disadvantage because the large number of small pixels
implies that, for a fixed number of pixels,
the largest pixels (and the average pixel diameter)
are bigger than in the 
other models.  
Figure \ref{fig:area} shows the distribution of area for the ECP and the equal 
latitude 3:6:3 scheme.  
While the pixel areas do vary in the 
equal latitude 3:6:3 scheme, they do so in a 
way which keeps the pixel distortion small.

{\bf Hierarchy:}
In regards to hierarchy, the best 
pixelization is the ECP, 
which in its simplest implementation can be reduced to just two pixels, 
the East and West hemispheres. 
The 3:6:3 models are nearly as good, having only twelve base pixels. Of the 
two, the equal latitude case is somewhat simpler and has the 
advantage that it can be subdivided into an ECP pixelization  
with $(3 \times 6)2^n$ pixels.    
Clearly, the poorest pixelization in this regard is the 12,116 
base pixelization, 
which only coarsens down to two degree patches, and so requires more 
information to specify uniquely. 

\section{Evaluating Pixelizations} 

The criteria we use to judge the various pixelizations are the speed and 
accuracy with which one can create and invert realistic sky maps.
Any reasonable pixelization should do well in finding the low multipoles
which do not vary appreciably over the pixels.  The true test is how 
well one can extract the power spectrum at high multipoles, corresponding 
to wavelengths on the sky approaching the dimensions of the pixel.

To evaluate different pixelizations, we create realizations assuming 
Gaussian statistics and a sample (CDM) power spectrum. The simulated multipole
coefficients are then exactly transformed into pixelized maps. 
We then perform an exact spherical transform of the pixelized maps to
attempt to recover the initial power spectrum. In this final step, 
errors occur in two ways. First, the pixelized maps contain only
$N_{tot}$ numbers: information regarding modes whose wavelength is 
shorter than the pixel size has been lost. Second, when one attempts
to correct the recovered multipole coefficients 
for the effect of pixel smoothing, one can do so at various levels 
of approximation (and speed), as will be discussed below. 

\subsection{General Considerations} 

Two maps must be considered, the true map of the temperature on the sky, 
and the map projected into the pixelization of the sky.  
The true 
continuous temperature can be expressed as 
\be 
T(\theta,\phi) = \sum_{l,m} a_{lm} Y_{lm}(\theta,\phi) 
\ee 
where the $a_{lm}$'s are variables that are Gaussianly distributed with 
zero mean and variance given by the power spectrum, 
$\langle |a_{lm}|^2 \rangle = C_l.$
This can be inverted using the orthogonality of the spherical harmonics to 
find 
\be 
a_{lm} = \int T(\theta,\phi) Y^*_{lm}(\theta,\phi) d\Omega.    
\ee
For our present purposes, the true map is the sky smoothed by the 
experimental window function.
The two point correlation function is
\bea 
 \langle T(\Omega)T(\Omega') \rangle  & = & 
\sum_{lm}C_l Y_{lm}(\Omega) Y^*_{lm}(\Omega') \nonumber \\ 
& = & {1 \over 4\pi} \sum_l (2l+1)C_l P_l(\cos\theta_{\Omega\Omega'}),
\eea
where $\theta_{\Omega\Omega'}$ is the angular separation of the two points. 

Unfortunately, the true map is not directly observable. 
Experiments will produce 
pixelized maps where the temperature in each pixel is the true temperature 
averaged over the area of the pixel.
The measured temperature in one of these 
pixels is given by 
\be 
T_P = \int W^P(\theta,\phi) T(\theta,\phi) d\Omega 
\ee 
where $W^P(\theta,\phi)$ is  the pixel window function for the P$^{th}$ pixel, 
equal to $1/A_P$ (where $A_P$ is the area of the pixel) within the pixel
and zero otherwise.  Thus, $T_P$ is the average of the temperature in the 
pixel.  Expanding the temperature in terms of spherical harmonics, we find 
\bea 
T_P & = & \int W^P(\theta,\phi) \sum_{l,m} a_{lm} Y_{lm}(\theta,\phi) d\Omega
\nonumber \\
 & = & \sum_{l,m} a_{lm} W^{P}_{lm} \eea
where $W^{P}_{lm}$ is defined by 
$ W^{P}_{lm} = \int d\Omega  W^P(\theta,\phi) Y_{lm}(\theta,\phi)$ and 
$ W^P(\theta,\phi) = \sum W^P_{lm} Y^*_{lm}(\theta,\phi).$ 
The vector 
$ W^{P}_{lm}$, $ P = 1,..N_{tot}$, is the pixelized analog of the continuous 
function $Y_{lm}(\theta,\phi)$.  

The pixelized analog to the two point correlation function is the pixel-pixel
correlation, 
\be 
\langle T_P T_Q \rangle = \sum_{lm} C_l W^P_{lm} W^{Q*}_{lm}. 
\ee 
In general, this cannot be further simplified because the pixelization 
breaks rotational invariance.   For the special case of white noise, 
$C_l = C_{WN}$ is constant, and the pixel-pixel correlation is simply
$C_{WN} \delta_{PQ}/A_P$. 

While the pixelized map is discrete, it is a well defined function on the 
sky and can itself be expanded in terms of orthogonal functions: 
\bea T^{pix}(\theta,\phi) & = & \sum_P T_P A_P W^P(\theta,\phi)
\nonumber  \\
& = & \sum_{l,m} a^{pix}_{lm} Y_{lm}(\theta,\phi). \eea 
This can be inverted to find that 
\bea a^{pix}_{lm} & = &\int T^{pix}(\theta,\phi) Y^*_{lm}(\theta,\phi) d\Omega
\nonumber \\
& = & \int \sum_P T_P A_P W^P(\theta,\phi) Y^*_{lm}(\theta,\phi) d\Omega 
\nonumber \\ 
& = &  \sum_P T_P A_P W^{P*}_{lm}.\eea
This is the pixelized analog of equation (3.2).  
In principle, all the pixelized multipoles, an infinite number 
of them, are well 
defined and could be calculated.  However, since there are only 
$N_{tot}$ independent pixel temperatures, only a finite number of the extracted 
$a_{lm}^{pix}$'s can be linearly independent, which we 
can take to be the 
lowest $N_{tot}$ $a^{pix}_{lm}$'s (i.e. with $l$ up to
$l= \sqrt{N_{tot}}-1$). 

The pixelized multipoles can be related to the true multipoles
by substituting the above expression for $T_P$, 
\be a^{pix}_{lm} = \sum_{l'm'} a_{l'm'} \sum_P A_P
W^{P*}_{lm} W_{l'm'}^P. 
\label{eqn:exact}
\ee
This equation tells us that the $a^{pix}_{lm}$ are special linear
combinations of the true underlying $a_{lm}$'s we are interested in.
A pixelized map can only provide a finite number of constraints 
on the infinite number of possible multipoles which might exist. 
We cannot recover the underlying $a_{lm}$'s 
without making an additional assumption, the simplest being that
they are all zero beyond some value of $l$. 
This is perhaps the most natural choice, because 
while all the multipoles might in principle 
be non-zero, the highest multipoles will eventually be suppressed due 
to the finite resolution of the antenna and the pixel size.  
(If higher multipoles 
are present, then these will mimic combinations of the lower multipoles and 
so will reduce the accuracy to which they can be extracted.)
With this assumption,
we can attempt to solve equation (\ref{eqn:exact})  for the true $a_{lm}$'s.

One possible way to solve equation (\ref{eqn:exact}) for the $a_{lm}$'s
exactly is by iteration, making a guess for the $a_{lm}$'s and 
then making a correction based on that guess. 
Consider the positive semi-definite function
\be f= \sum_{P} A_P T_P^2({\rm error})
\label{eqn:func}
\ee
where $T_P({\rm error})$ is the temperature difference between the 
measured temperature on the $P$'th pixel and the temperature one would 
get given some guess for the $a_{lm}$'s, 
\be T_P({\rm error}) = T_P({\rm guess})- T_P({\rm true}).
\ee
The scheme to solve 
(\ref{eqn:exact}) is then just to minimize the function $f$ in (\ref{eqn:func}),
which can be straightforwardly calculated for any conjectured 
set of $a_{lm}$'s using the fast Fourier transforms
described below. The minimization of 
$f$ can be done iteratively in a very simple manner, 
and we report on results from this method below. 

One danger in making this inversion is that the multipoles we have 
chosen to extract might not 
fully span all the degrees of freedom of temperature map.  
If this were the case, the matrix in equation (\ref{eqn:exact}) is singular 
and cannot be inverted.
For example, two apparently orthogonal functions might 
appear identical when they are pixelized, making a unique inversion 
for the $a_{lm}$'s impossible.
In reality, this degeneracy of functions will only be approximate, 
but it will make solving for the underlying $a_{lm}$'s difficult. 
To minimize these dangers, it makes sense for the ideal pixelization to 
represent all the multipoles we have chosen to solve for
(those below $l \le \sqrt{N_{tot}}-1$) 
as orthogonally as 
possible.

While it is important that in principle a 
scheme exists for solving (\ref{eqn:exact}) exactly for the $a_{lm}$'s 
(modulo the necessary assumption that they vanish for high $l$),
in practice one may be more interested in a fast scheme for
recovering the $a_{lm}$'s approximately. 
When the spherical harmonics do not vary much across the 
pixels (i.e., for low $l$), then the orthogonality of the spherical harmonics, 
$\int d\Omega Y_{lm}Y^*_{l'm'} = \delta_{mm'}\delta_{ll'} $
implies that 
\be \sum_P A_P W^{P*}_{lm} W_{l'm'}^P \simeq N_{lm} \delta_{ll'}\delta_{mm'}, 
\ee 
where, $N_{lm} \equiv \sum_P |W^{P}_{lm}|^2 A_P $.
In this case, the inversion can be found to be 
\be 
a^{est}_{lm} \simeq {\sum_P T_P A^P W^{P*}_{lm} \over N_{lm} }. 
\label{eqn:approx}
 \ee
This relation naturally breaks down at high multipoles, when the number of 
multipoles to be extracted approaches  the number of pixels.

One can estimate the power spectrum from these extracted multipole moments. 
The naive estimate is simply, 
\be 
C_l^{est} \equiv {1 \over (2l+1)} \sum_m |a^{est}_{lm}|^2 . 
\ee 
Unfortunately, this estimator is biased and tends to overestimate 
the power spectrum, especially at high multipoles.  This is easy to 
understand.  Suppose one has a single mode on the sky, which is 
then pixelized. 
The lack of orthogonality of the modes projected on 
the pixelization implies that when inverting the pixelization, one would 
recover the original mode, plus an admixture of 
small contributions to other modes.  
The latter tend to increase the extracted power.  
One can construct an unbiased estimator from $C_l^{est}$ 
(or from $C_l^{pix}$.)  
It can be shown that  
\be 
\langle C_l^{est} \rangle = {\cal M}_{ll'} C_{l'} 
\ee 
where, 
\be 
{\cal M}_{ll'} \equiv {1 \over (2l+1)} \sum_{mm'} {1 \over N_{lm}^2}
\sum_P A_P W^{P*}_{lm} W_{l'm'}^P \sum_Q A_Q W^{Q}_{lm} W_{l'm'}^{Q*}. 
\ee 
This matrix is a function of the pixelization alone, so need be calculated 
only once.  Given this matrix, an unbiased 
estimator of the power spectrum is 
\be
C_l^{unbiased} = {\cal M}^{-1}_{ll'} C^{est}_{l'}. 
\label{eqn:unbias}
\ee
This removes the systematic effect of leakage from other multipoles 
and is quick to calculate.
However, it ignores off-diagonal correlations between the 
$a_{lm}^{est}$'s, and so is not the best possible estimator of 
the power spectrum. 

\subsection{Azimuthal Pixelizations} 

For azimuthal pixelizations, it is helpful to replace the general pixel index 
$P$, with two indices $p,q$ where the former represents the row number 
and the latter identifies the pixel within that row.  
Each row has $N_p$ identical pixels placed periodically in $\phi$. 
We can expand the spherical harmonics as 
\be Y_{lm}(\theta,\phi) = \lambda_{l}^m(\cos \theta) e^{i m\phi}, \ee
where $\lambda_{l}^m(x)$ are the normalized Legendre polynomials, 
defined in terms of the associated Legendre polynomials by 
\be 
\lambda_{l}^m(x) = \left( {(2l+1) \over 4 \pi }{(l-m)! \over (l+m)!} 
                \right)^{1/2} P_{l}^m(x),
\ee
following the notation of Muciaccia, Natoli and Vittorio (MNV, 1997).  

The window functions in azimuthal pixelizations can be easily calculated. 
\bea 
W^{pq}_{lm} &=& \int d\Omega  W^{pq}(\theta,\phi) Y_{lm}(\theta,\phi) 
\nonumber \\
&=& {1 \over A_p}\int^{2\pi q/N_p}_{2\pi(p-1)/N_q} e^{i m\phi} d\phi
\int^{\theta_{p+1}}_{\theta_p}
\lambda^m_l(\cos \theta) \sin\theta d\theta \nonumber \\
&=& e^{i mq2\pi/N_p}W^{p}_{lm} \eea 
where we have defined the $q$ independent part of the window function as, 
\be W^{p}_{lm} = e^{i m\pi/N_p}{\sin(\pi m/N_p)\over (\pi m/N_p)} 
{1 \over \Delta x_p}\int^{\theta_{p+1}}_{\theta_p}
\lambda^m_l(\cos \theta) \sin\theta d\theta,   
\ee
and $\Delta x_p = \cos(\theta_{p+1}) - \cos(\theta_p)$.
The final integral can be evaluated very quickly in a recursive fashion, 
comparable to the time required to evaluate the Legendre functions themselves. 
(See appendix A.)  The ability to exactly integrate over the pixels quickly is 
an important advantage of the igloo class of tilings. 

It is the simple dependence on $q$ of the window functions 
that allow the creation and inversion of maps to be evaluated using Fast 
Fourier Transforms. 
To take advantage of this in creating maps, it is useful to rewrite the 
expression for the pixel temperature, reordering the sums in a suggestive way
(again adapting the notation of MNV):
\bea 
T_{pq} & = &  \sum_{l=0}^{l_{max}} \sum_{m = -l}^l a_{lm} W^{p}_{lm} 
e^{i mq2\pi/N_p}  \nonumber \\ 
& = & \sum_{m = -l_{max}}^{l_{max}} \sum_{l=|m|}^{l_{max}} a_{lm} W^{p}_{lm} 
e^{i mq2\pi/N_p} \nonumber  \\
& = & \sum_{m = -l_{max}}^{l_{max}} b_m(\theta_p) e^{i mq2\pi/N_p}.\eea
Here, 
\be 
b_m(\theta_p) = \sum_{l=|m|}^{l_{max}} a_{lm} W^{p}_{lm} = 
\sum_{q}T_{pq} e^{-i mq2\pi/N_p}, 
\ee 
and the transformation between $b_m(\theta_p)$ and $T_{pq}$ can 
be made with a FFT. 
However, since there are only a limited number of pixels, only the frequencies 
$-N_p/2 \le m \le N_p/2$ can be represented on the lattice.
When the number of modes exceeds $N_p/2$, the modes appear on the lattice
like modes with frequency $mod(m,N_p)$ and their power must be aliased into 
these lower frequencies.  

Since the northern and southern hemispheres are tiled in the same way, we can 
use this symmetry to save a factor of two in computing time \cite{moh97}.
The associated 
Legendre polynomials are even or odd functions across the equator,
depending on whether
$l+m$ is an even or odd number.
We can break the $b_m(\theta_p)$ functions into even and odd pieces and evaluate
them in only the northern hemisphere, and find the full function by
\bea
b_m(\theta_p,{\rm north}) &=& b_m(\theta_p, {\rm even}) + b_m(\theta_p, {\rm odd})  \\
b_m(\theta_p,{\rm south}) &=& b_m(\theta_p, {\rm even}) - b_m(\theta_p, {\rm odd}).
\eea

The inversion of the maps can be performed similarly,
\bea 
 a_{lm} & = & {1 \over  N_{lm}} \sum_{pq} T_{pq} A^p W^{pq*}_{lm} \nonumber  \\
& = & {1 \over  N_{lm}} \sum_{p=1}^{n_{rows}} A^p W^{p*}_{lm} 
\sum_{q}T_{pq} e^{-i mq2\pi/N_p} \nonumber \\
& = & {1 \over  N_{lm}} \sum_{p=1}^{n_{rows}} b_m(\theta_p) A^p W^{p*}_{lm}.
\eea
Similarly, care must be taken in the inversion to map the power from the lower
modes to the higher ones.  

It is the effect of the higher modes on the pixelization mimicking the lower 
ones which is primarily responsible for making the exact inversion of the 
map impossible.  
In the continuous case, the orthogonality of the spherical harmonic functions 
is enforced in two ways: first, the azimuthal functions 
($e^{i m\phi}$) are orthogonal
for $m \ne m'$ and second, when $m = m'$ (and only then), 
the Associated Legendre functions 
are orthogonal for $l \ne l'$.  In the pixelized case, this remains approximately 
true when $m,m' \le N_p/2$, but this breaks down for larger modes, so that modes
with $m \ne m'$ are not necessarily orthogonal.  Then the lack of orthogonality 
for the Legendre polynomials means these functions can appear to be the same.

To minimize this effect, it is helpful to have as many pixels per azimuthal 
row as possible.  For the equal area models we consider, each pixel has an 
area of $4\pi/N_{tot}$, with edges of length $\sqrt{4\pi/N_{tot}}$.  Thus there are 
$\sqrt{\pi N_{tot}}$ pixels along the equator, and the problem of orthogonality 
becomes a serious problem near the Nyquist frequency at 
the equator, $l \simeq N_{equator}/2 = \sqrt{\pi N_{tot}}/2$.  
This is close to the 
maximum we could reasonably hope for given the finite number of 
pixels, $l \simeq \sqrt{N_{tot}}$.

The normalization of each mode can be found quickly from the expression, 
\be 
N_{lm} = \sum_{p=1}^{n_{rows}} A_p W^{p*}_{lm} W_{lm}^p N_p. 
\ee
For round pixels, the normalization is independent of $m$ and this 
is usually assumed for the sake of simplicity.  In a 
more realistic pixelization, however, there can be large variations for 
different $m$'s, especially at high multipoles.  Thus, the ability to 
calculate these individual normalizations is a great advantage in 
making an inversion.  

\section{Results} 

We begin by showing the average normalization, 
$\sum_m N_{lm}/(2l+1)$,  for the various pixelizations. 
This illustrates the average suppression of modes by the 
pixelization and is equivalent to its window function. 
The mean window functions are virtually identical for all the pixelizations 
with the exception of the ECP which falls off slightly more quickly. 
The shape of these curves reflects primarily the effective area of the pixels, 
and while the ECP has the same average area, its skewed 
distribution of shapes makes 
them dominated by the larger pixels.  

\begin{figure}[htbp]
\centerline{\psfig{file=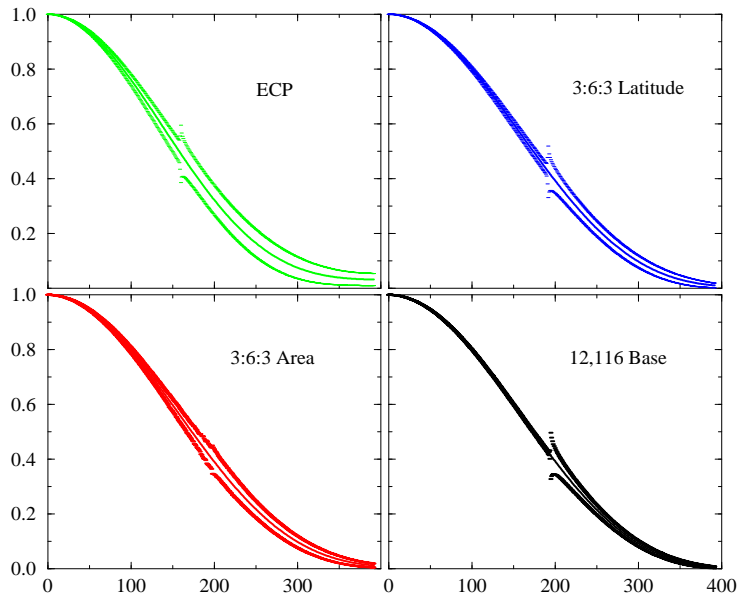,width=3.9in}}
\caption{ Figure of $N_{lm}$ averaged over $m$ as a function of $l$ for each 
pixelization.  The error bars indicate the variance of $N_{lm}$ 
for the different 
values of $m$.  (Based on roughly 49,000 pixels.)
}
\label{fig:window}
\end{figure}

A more pronounced difference appears in the variance of the window functions 
for a fixed $l$, as indicated by the error bars.  This is correlated with the 
average amount of distortion of the pixelization, with the 
ECP having the largest 
variance and the 12,116 base the least.  
In addition, a feature appears in the variance of the window 
functions precisely 
at $l_{Nyquist}$, when particular modes can be either in phase or out of 
phase with the pixelization.  

To minimize aliasing of the higher multipoles, the ideal window function 
would be zero above a given cutoff in $l$ ($= \sqrt{N_{tot}}$,) so that the
pixelization would have no response to the higher modes. 
However, we have been assuming the pixels have a sharp cutoff in 
real space, which makes aliasing from higher $l$ modes inevitable. 
For both the temperature map and its Fourier transform to be as local 
as possible, it could be advantageous for the window function to be 
effectively Gaussian, 
falling off both at large distances and at high multipoles.  
However, such a scheme would make eliminating foregrounds more difficult.

\begin{figure}[htbp]
\centerline{\psfig{file=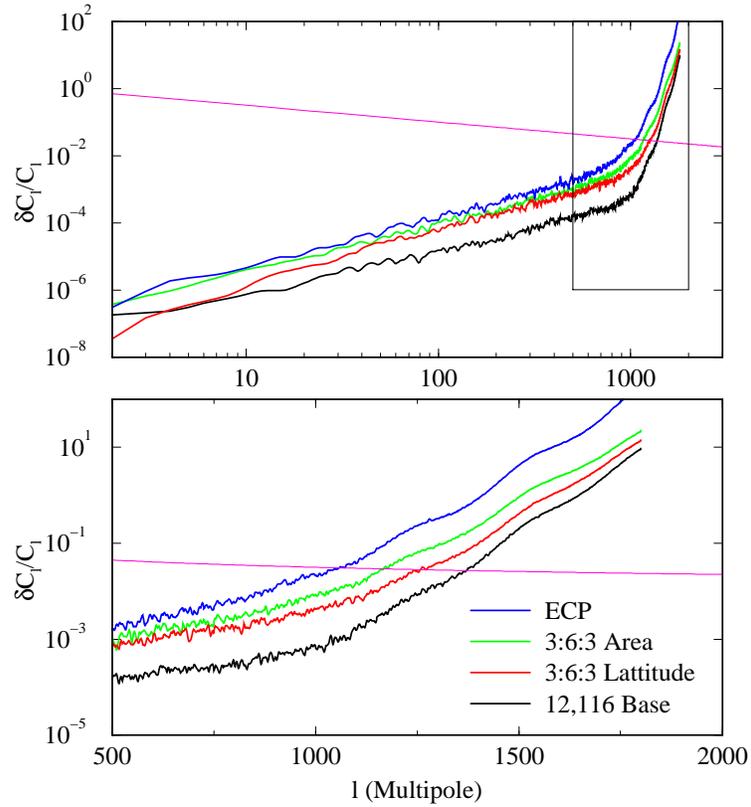,width=3.9in}}
\caption{ Figure of the errors in extraction of the 
$C_l$ for the various pixelizations
with roughly three million pixels in each.   
The boxed section is enlarged in the lower plot. 
The cosmic variance ($\delta C_l \sim C_l/\sqrt{l}$)
error level is shown the descending line, while the highest rising line 
shows  the effect of not using the exact window functions. The results are 
the average of five runs each, and are smoothed for clarity.
}
\label{fig:delta}
\end{figure}

How well can we recover an input power spectrum? 
We first consider inverting using the fast, approximate approach described 
in equation (\ref{eqn:approx}).
Figure \ref{fig:delta} shows the errors in inverting the power spectra 
using that technique for the four models we are considering. 
Here and below, we plot the fractional error in the estimated multipole, 
$(C_l^{est} - C_l^{made})/C_l^{made}$ and take the absolute value 
when plotting on a log scale. 
For this figure we have assumed a standard CDM power spectrum with a 
smoothing of $6'$.  We compare the various pixelizations with 
approximately three million pixels in each.
The accuracy seems to follow a power law ($\propto l^2$) at small multipoles 
and then rises quickly at higher multipoles, 
when 
the orthogonality of the modes breaks down and the power spectrum is 
systematically overestimated.  
At these multipoles, aliasing due to the other multipoles cannot be ignored.

As one might expect, due to the larger average diameter, 
the ECP pixelization performs worst 
of the models we consider, 
for fixed number of pixels. 
The best performing model is the 12,116 base 
pixel model, which has been shown to have the least pixel distortions and 
has exactly equal area.  
It is able to invert the power spectrum an order of magnitude 
better than the other models considered for most of the multipoles. 
In the middle range lie the 3:6:3 models, 
which have a smaller diameter than the ECP
model, but are more distorted than the large base pixel model.  
Here, the equal latitude model 
works somewhat better than the equal area models, 
implying that the diameter of the pixels 
is a key factor evaluating the pixelizations. 

For comparison, we also include a rough measure of the level of cosmic 
variance which occurs because we can only measure a finite number of 
multipoles for a given $l$.  This sample variance drops roughly as 
$\delta C_l \sim C_l/\sqrt{l}$.  At low $l$, the uncertainties in 
extracting the multipoles are far outweighed by the sample variance.
These converge as one goes to higher and higher $l$. 
While the errors from the pixelization can be generally be made 
much smaller than the cosmic variance, they can not be ignored if 
they are systematic in nature. 

\begin{figure}[htbp]
\centerline{\psfig{file=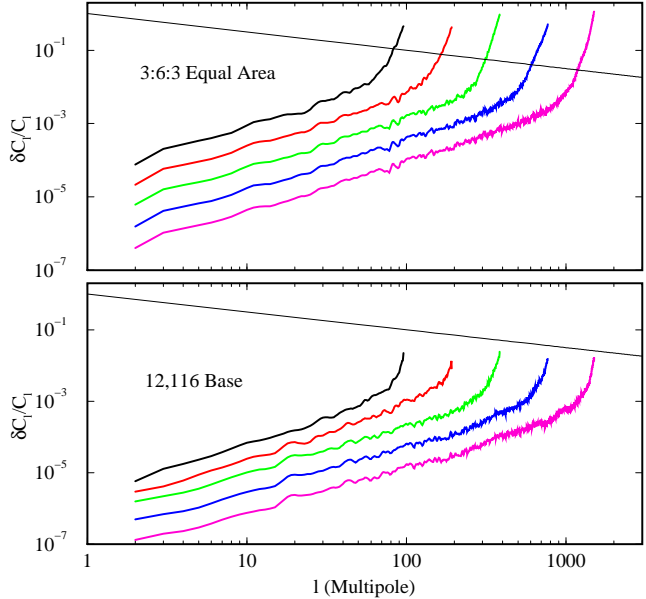,width=3.9in}}
\caption{ The scaling of the accuracy of the inversion for 
different resolutions for two pixelizations. 
Note that the features that 
seem to occur at particular multipoles are due to the particular seeds used 
to create the maps rather than
features inherent in the pixelization.
Shown are results for pixelizations with roughly $12,000 \times (1, 
4, 16, 64, 256)$ pixels.
}
\label{fig:scale}
\end{figure}

It important to understand how these conclusions change when the resolution
increases, keeping the same base number of pixels.  
Figure \ref{fig:scale} shows 
two of the pixelizations for a range of resolutions, from $2^\circ$ to 7'.  
To good approximation, the effect of increasing the number of pixels 
by four is to 
shift the curves to higher $l$ by a factor of two.  
(The residual differences between the curves is likely 
to be due to the varying shape of the input power spectrum.)
At low $l$, increasing the number of pixels by a factor of four decreases the 
errors in the extracted power by approximately the same factor.   

\begin{figure}[htbp]
\centerline{\psfig{file=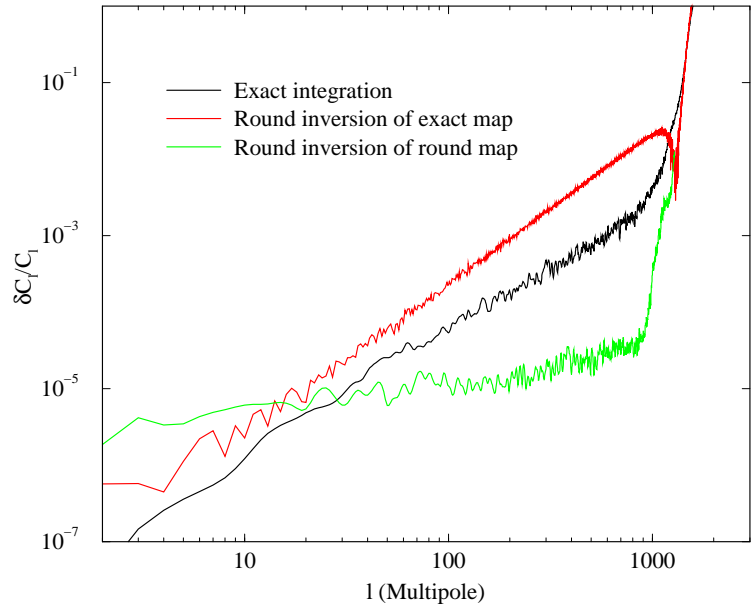,width=3.9in}}
\caption{ We show the effects of approximating the pixels as 
round tophat pixels with the same average area.  If one creates 
and inverts the maps using this assumption, one finds very good 
accuracy compared to the exact results.  These results are deceptive, however,
because they ignore the true pixel shapes. 
If one creates the maps by integrating over the true pixel shapes, and
inverts assuming the pixels were round tophats, the accuracy is much worse 
than the exact results.
}
\label{fig:point}
\end{figure}

Our results are qualitatively 
different from those found by MNV, who found the percentage errors were roughly 
constant as a function of 
$l$. 
This is likely because they were creating and inverting the maps 
using the same assumptions, evaluating 
the spherical harmonics in the center of the pixels, rather than 
integrating over the pixels as we do here.  
In this approach, pixel smoothing can be approximated 
by assuming the pixels are uniform round tophats with the same area of the 
the true pixels.
However, creating the maps in this way 
effectively ignores the true pixel shapes and leads to 
a sizable underestimation of the errors arising from the pixelization.
Figure \ref{fig:point} shows the effect of inverting such a realistic 
map by assuming round pixels.  
One can see that the effect of this is 
to greatly reduce the accuracy of the inversion,
even for the lowest multipoles.  The systematic error introduced by 
using an inaccurate window function appears to dominate the statistical errors.

When testing the quality of their inversions, 
previous analyses have 
assumed that the power spectrum of the initial maps cuts off precisely 
at the Nyquist frequency on the equator, $l_{Nyquist} = N_{equator}/2$. 
Thus, they have only put in the power that they could extract.
However, this is not a realistic assumption -- the higher multipole modes 
exist and 
must be included to understand the accuracy of the 
extracted power spectrum. 
Though these modes tend to be damped by experimental and pixel smoothing, 
even when the pixel size is much smaller than the beam size the spectrum 
falls off smoothly rather than sharply.  Thus, the power above any given 
cutoff is always comparable to that just below it and cannot be ignored.  
The presence of these higher modes can adversely affect 
the inversion because they can mimic the lower modes on the pixelization.

\begin{figure}[htbp]
\centerline{\psfig{file=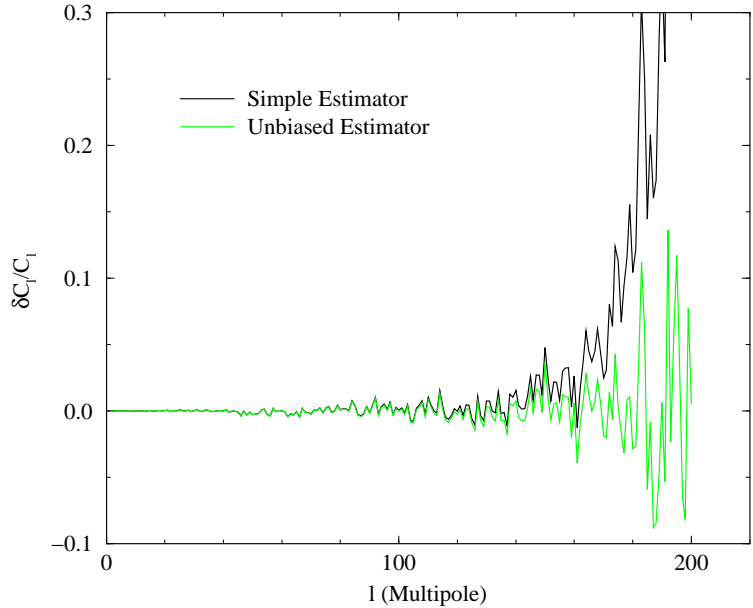,width=3.9in}}
\caption{The simple estimator for the power spectrum is 
biased by the presence of other modes, which leads to systematically 
overestimating the power.  Corrected for this bias greatly 
reduces the error at high $l$. 
(Based on $N_{tot} = 49,152$ pixels, with $l_{Nyquist} = 192$.)
}
\label{fig:unbias}
\end{figure}

The presence of the higher modes can be corrected 
for, in part, by using the unbiased estimator defined in 
equation (\ref{eqn:unbias}).  Figure \ref{fig:unbias} shows 
the errors in the recovered power spectra with and without 
correcting for this bias.   Without the correction, the power spectrum is 
greatly overestimated at high frequencies.  
With the correction, the power spectrum can be recovered even beyond 
$l_{Nyquist}$, though with decreasing accuracy.

We have been comparing models assuming they have nearly the 
same numbers of pixels. 
Rather than using the number of pixels as 
our criterion,
it might be fairer to compare pixelizations for a 
given speed, or equivalently a given number of azimuthal rows.  
For example, the ECP pixelization has $\sim$ 20\% fewer rows than 
the other models  we considered and would perform much better  
if it had the 
same number of rows, when it would have 50\% more pixels. 
This would improve its accuracy, making it comparable to the 
12,116 base model. 
It is worth considering whether for a given pixelization one should 
increase the number of pixels per row, which would improve the orthogonality of 
spherical harmonics without adding much to the computation time.  
(Though, some it would not improve much, for example the $m=0$ modes.) 
If one chooses 
to make the pixels more narrow, however, 
it would be better to do it in a uniform way for all of the pixels, 
rather than just at the poles, as the ECP pixelization does.   
 
Finally, we have also attempted to invert equation  (\ref{eqn:exact})
directly rather than only approximately.
We constructed simulated skies in which we set the $a_{lm}$'s equal to 
zero beyond some cutoff value $l_c$ and used an iterative scheme  
described in \cite{NR} to find the minimum of the function $f$, defined 
in equation (\ref{eqn:func}), taking advantage of the sparseness of the matrix. 
One difficulty in an exact inversion is that 
as $l_c$ approaches $\sqrt{N_{tot}}$, the number of directions 
in which $f$ is flat grows. That is, some particular linear 
combinations of $a_{lm}$'s have very little impact on 
the pixelized temperatures: the pixelization effectively washes
out these combinations.  These multipoles are thus very difficult 
to extract.

\begin{figure}[htbp]
\centerline{\psfig{file=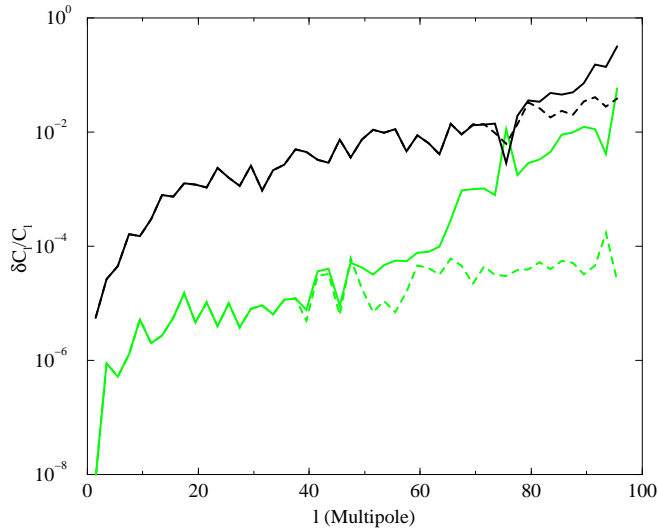,width=3.9in}}
\caption{  A comparison of the accuracy of an approximate fast inversion (the 
darker lines) to 
an iterative method of inversion (the lighter lines.)  The accuracy is limited 
by a few modes which are difficult to extract accurately.  When these 
modes are excluded, the accuracy improves dramatically (dashed lines.)
Here, approximately 10\% of the multipoles have been excluded.  
Note, we plot the accuracy of the recovery of the remaining $a_{lm}$'s.
}
\label{fig:correct}
\end{figure}

To overcome this problem, we conduct 
a number of trial simulations and minimizations, and 
compute the fractional error $\langle|\delta a_{lm}|^2\rangle/C_l$ for
each $l$ and $m$. Modes where this fractional error
exceeds some threshold are those which 
are hardest to recover. These modes are 
a function of the pixelization and are the same from run to run. 
The most naive prescription is then
to just ignore the recovered $a_{lm}$'s for these `hard' values
of $l$ and $m$. Figure \ref{fig:correct} shows results for $l_c=96$, 
$N_{tot}=12,288$, where the number of multipoles which one is ignoring is
of the order of ten per cent of the total (most
of these are at $l$ very close to  $l_c$). The errors in 
the recovered $C_l$'s shown are for a run independent from 
those which determined which modes were to be excluded. 

This graph demonstrates that there is much recoverable information
available in the pixelized maps about modes up to $l \sim \sqrt{N_{tot}}$, 
at the cost of increased computational effort. 
However, we have performed this inversion using a very simple scheme, and 
there may well be more efficient ways of recovering it. 
In particular it may be worthwhile to consider using two 
pixelizations in which the `poles' lie along orthogonal directions. 
The underlying $a_{lm}$'s which were hardest to recover from
one pixelization would be very different from those which 
were hardest to recover from the other. So replacing the function $f$ in
(\ref{eqn:func}) by the sum of the two functions $f$ for each pixelization would
likely remove some of the flat directions.

\section{Conclusions}

We have shown that simple igloo pixelizations
have much to commend them.
They can easily be made 
hierarchical with pixels that are quite uniform. 
More importantly, they are naturally 
azimuthal, which speeds the calculation of spherical harmonics and
allows for the exact calculation of the effect of the pixel shape. 
Including the effects of the pixel shapes is essential 
for the accurate extraction of the power spectrum 
without greatly over sampling 
the sky, and for understanding the systematic errors which 
can arise.  

We should emphasize that in the limit of an infinite number of pixels, 
all the pixelizations are equivalent.   In practice, the beam 
will be oversampled, which will help reduce the effects 
of pixelization for the multipoles of greatest interest.  
However, computational pressure will drive the number of pixels down, 
and remnants of the pixelization will exist to some degree.  
Igloo pixelizations provide an exact way of correcting for them. 

In deciding what particular pixelization to use, a
tradeoff must be made between the simplicity inherent in having few base 
pixels and the advantages of having pixels 
with the smallest amount of distortion possible.
We have shown that having many base pixels has advantages in making the 
spherical harmonic representations on the pixelizations orthogonal, 
which minimizes errors in extracting power spectra.
However, these factors must be weighed against the practical advantages 
having a simple and hierarchical scheme. 

Of the models we have considered here, perhaps the best compromise 
between these considerations 
is the 3:6:3 equal latitude model. 
It has only twelve base pixels and can easily be created from an ECP 
pixelization. 
While it is only approximately equal area, we have found that this is actually 
an advantage because it allows us to prevent the average pixels from becoming 
more distorted. 
If exactly equal area models are favored, these can 
also be implemented using a 3:6:3 base igloo 
model.   

Finally, we emphasize that 
there are many issues which we have not addressed here.  
For simplicity, we have assumed full sky coverage, 
though in practice the galaxy foreground will have 
to be removed. 
We also have focused 
on an ideal situation with no experimental noise, while realistically this will 
be large and will vary across the sky. 
These issues are crucial to eventually understanding the 
microwave background, but we have put them aside here 
to isolate the effects that arise from the process of the pixelization alone.

We wish to thank Richard Battye, Kris Gorski, 
Steven Gratton, David Spergel and Ned Wright 
for useful conversations, as well as members of the Cambridge Planck 
Analysis Center.  R.C. acknowledges support from a PPARC Advanced Fellowship.  

\appendix
\section{Fast Integration of Legendre Polynomials}

In calculating the true window function of a particular pixelization, it 
is useful to have an algorithm for the quick calculation of $\int P_{l}^mdx$ 
and its normalized counterpart, $\int \lambda_{l}^mdx.$  Just as for the 
calculation of the $P_{l}^m$'s themselves, they can be calculated by 
means of a recurrence relation.  

Two of the recurrence relations that hold for the Associated 
Legendre polynomials are \cite{GR}
\bea
(1-x^2) {d P_{l}^m(x) \over dx} & = &(l+1)xP_{l}^m(x) - (l-m+1)P_{l+1}^m(x), 
 \\
(2l+1)xP_{l}^m(x) &=& (l-m+1)P_{l+1}^m(x) + (l+m) P_{l-1}^m(x). \eea 
Numerically, 
equation (A2) is generally used recursively to evaluate the 
Legendre polynomials, beginning with $P_{m}^m(x)$ and $P_{m+1}^m(x)$ 
\cite{NR}.

Integrating the first relation, one finds 
\be
(1-x^2) P_{l}^m(x) = (l-1)\int xP_{l}^m(x) dx - (l-m+1)\int P_{l+1}^m(x)dx.
\ee
Now substituting in the second relation, we find the recursion relation 
for the Legendre polynomials 
\be 
\int P_{l+1}^m(x)dx = {(l-1) \over (l+2)}{(l+m) \over (l-m+1)} 
\int P_{l-1}^m(x)dx
- {2l+1 \over (l+2)(l-m+1)} (1-x^2) P_{l}^m(x).
\ee 
In terms of the normalized polynomials, the recursion relation is, 
\be 
\int \lambda_{l+1}^m(x)dx = \left({2l+3 \over (l+m+1)(l-m+1)}\right)^{1/2} 
{1 \over l+2} \left( {(l-1)(l^2-m^2)^{1/2} \over (2l-1)^{1/2}}
\int \lambda_{l-1}^m(x)dx
- (2l+1)^{1/2} (1-x^2) \lambda_{l}^m(x) \right).
\ee 

This recursion requires the initial values, $\int \lambda_{m}^m(x)dx$ and 
$\int \lambda_{m+1}^m(x)dx$.   The first of these can be found recursively, 
working up from $m=0,1$.
However, this recursion becomes unstable when $\sin^m(\theta)$ is very small. 
When this is the case, one can rewrite the recursion as an expansion in 
powers of $1 / m\cos^2\theta$, which quickly converges. 

The second integral is given simply by the identity, 
$\int \lambda_{m+1}^m(x)dx = -(1-x^2) \lambda_{m}^m (2m+3)^{1/2}/(m+2)$. 
To avoid round-off error, we perform the recursion for the integrals
over the range of interest $\int^{x_1}_{x_2}\lambda_{l}^m(x)dx$, rather than 
doing them separately 
and taking the difference (i.e. 
$\int^{x_1}_0\lambda_{l}^m(x)dx - \int^{x_2}_0\lambda_{l}^m(x)dx$.)

\section{Pixelization as a Projection Operator}

Pixelization projects the sky, which 
has infinite degrees of freedom parameterized by its $a_{lm}$'s, 
into a subspace of $N_{tot}$ dimensions, parameterized by the pixel 
temperatures, $T_P$.  Thus, one can think of the operation which takes 
the true map to the pixelized map as a projection operator.  

In real space this is self evident -- pixelization 
projects the multi-dimensional temperature map over a pixel into a 
one dimensional space parameterized by 
its averaged temperature.  
In multipole space, this interpretation is less obvious. 
From equation
(\ref{eqn:exact}), we can define the projection operator 
in multipole space to be 
\be 
{\cal P}_{lml'm'} \equiv \sum_P A_P W^P_{lm}W^{P*}_{l'm'} 
\ee 
so that 
\be 
{\cal P}_{lml'm'} a_{lm} = a_{l'm'}^{pix} 
\ee 
where the repeated indices are summed over. 

Projection operators have the property that ${\cal P}^2 = {\cal P}$. 
One can show this for the operator defined above 
using the completeness of the spherical harmonics: 
\be 
\sum_{lm} Y_{lm}(\Omega) Y^*_{lm}(\Omega') = \delta(\Omega - \Omega'). 
\ee
Multiplying both sides by $W^P(\Omega) W^Q(\Omega')$ and integrating over both
solid angles, one finds the pixelized analog 
\be 
\sum_{lm} W^P_{lm} W^{Q*}_{lm}  = {1 \over A_P} \delta_{PQ}. 
\ee 
Thus, 
\bea 
{\cal P}_{lml'm'} {\cal P}_{l'm'l''m''} & = & 
\sum_P A_P W^P_{lm}W^{P*}_{l'm'} \sum_Q A_Q W^Q_{l'm'}W^{Q*}_{l''m''}
\nonumber \\
& = & \sum_P A_P W^P_{lm} \sum_Q A_Q W^{Q*}_{l''m''} {1 \over A_P} \delta_{PQ}
\nonumber \\ & = & \sum_P A_P W^P_{lm} W^{P*}_{l''m''} = {\cal P}_{lml''m''}
\eea 
In real space, this  
is equivalent to saying that pixelizing a map that has already been pixelized 
yields the same map.

\end{document}